\documentclass[reprint,amsmath,amssymb,prb,floatfix,twocolumn,superscriptaddress]{revtex4-1}
\usepackage[table]{xcolor}
\usepackage{graphicx}
\usepackage{dcolumn}
\usepackage{times}
\usepackage{colordvi,epsf}
\usepackage[normalem]{ulem}
\usepackage{soul}

\usepackage{graphicx}
\usepackage{amsmath}
\usepackage{amssymb}
\usepackage{bm}
\usepackage{pgf}
\usepackage{color}
\usepackage{amsfonts}
\usepackage{hyperref}
\usepackage{tcolorbox}
\usepackage[bottom]{footmisc}

\begin{document}

\title{Machine learning magnetic parameters from spin configurations}
\author{Dingchen Wang}
\altaffiliation{Dingchen Wang\& Songrui Wei Contributed equally to this work}
\affiliation{MOE Key Laboratory for Nonequilibrium Synthesis and Modulation of Condensed Matter, School of Science, State Key Laboratory for Mechanical Behavior of Materials, Xi’an Jiaotong University, Xi’an 710049, China}


\author{Songrui Wei}
\altaffiliation{Dingchen Wang\& Songrui Wei Contributed equally to this work}
\affiliation{Key Laboratory of Optoelectronic Devices and Systems of Ministry of Education and Guangdong Province, College of Optoelectronic Engineering, Shenzhen University, Shenzhen 518060, China}

\author{Anran Yuan}
\affiliation{
	Key Laboratory of Intelligent Perception and Image Understanding of Ministry of Education, International Research Center for Intelligent Perception and Computation, Joint International Research Laboratory of Intelligent Perception and Computation, School of Artificial Intelligence, Xidian University, Xi’an 710071, China}

\author{Fanghua Tian}
\author{Kaiyan Cao}
\author{Qizhong Zhao}
\author{Chao Zhou}

\author{Yin Zhang}
\author{Xiaoping Song}
\author{Dezhen Xue}
\email{xuedezhen@mail.xjtu.edu.cn}
\affiliation{MOE Key Laboratory for Nonequilibrium Synthesis and Modulation of Condensed Matter, School of Science, State Key Laboratory for Mechanical Behavior of Materials, Xi’an Jiaotong University, Xi’an 710049, China}

\author{Sen Yang}
\email{yang.sen@mail.xjtu.edu.cn}
\affiliation{MOE Key Laboratory for Nonequilibrium Synthesis and Modulation of Condensed Matter, School of Science, State Key Laboratory for Mechanical Behavior of Materials, Xi’an Jiaotong University, Xi’an 710049, China}

\begin{abstract}
Hamiltonian parameters estimation is crucial in condensed matter physics, but time and cost consuming in terms of resources used.
%
%
High-resolution images provide detailed information of underlying physics, which can serve as input to machine learning (ML) algorithms to extract Hamiltonian parameters.
%
%
Here, we provide a protocol for Hamiltonian parameters estimation based on a ML architecture, which is trained on merely a small amount of simulated images and directly applied to a particular experimental observation.
With the single experimental observation as the only input, we are able to estimate all the key parameters simultaneously, which are employed to predict the materials properties. 
%
%
Our data augmentation method allows us to rapidly construct a convolutional neural network (\emph{CNN}) based on several images simulated under a particular experimental condition.
Therefore, we can deploy such a \emph{CNN} for any new experimental observation to estimate its Hamiltonian parameters efficiently.
%
%
We demonstrate the success of the estimation by reproducing the same spin configuration with the experimental one and predicting the coercive field, the saturation field and even the volume of the experiment sample accurately.
%
%
%
%
%
%
%
Our approach paves a way to achieve a stable and efficient parameters estimation.
\end{abstract}

\maketitle


\section*{I. INTRODUCTION}
Theoretical models describe the underlying physics of a given physical system and are able to understand and predict properties of a particular system if the model parameters are estimated appropriately. \cite{Cartwright1983How}
A typical example is the micro-magnetic model which evolves the spin configurations to the stable state according to the magnetic Hamiltonian. \cite{Burgess2013Quantitative}
%
Usually, the Hamiltonian considered include several terms of energy. 
The magnetic parameter exists in the Heisenberg exchange energy which tries to align neighboring spins, the Dzyaloshinskii-Moriya interaction which favors the canting of neighboring spins, and the Zeeman energy which is due to the external magnetic field and tries to align the spin with the field.
The strength of these contributions are controlled by parameters such as the Heisenberg exchange stiffness ($A_{ex}$), the Dzyaloshinskii-Moriya strength ($DMI$) and the saturation magnetization ($M_{sat}$), respectively. \cite{leliaert2018fast}
If the three key parameters are estimated properly,
%
%
%
%
%
many static and dynamical phenomena of artificial spin ice, Skyrmion, spin-waves and spintronics can be reproduced and predicted. \cite{Thiaville2002Domain}
Thus the Hamiltonian parameters estimation is essential in predicting and understanding properties of specific systems. \cite{Sampaio2013Nucleation, Zhou2015Dynamically, Farhan2013Exploring, Ramirez1999Zero, Wang2006Artificial, Ladak2010Direct, Tchernyshyov2010Magnetic}

However, since the estimation requires detailed control and measurements, as well as extensive postprocessing of the measured data, it is highly time and cost consuming \cite{zhang2014quantum,eyrich2012exchange,Buford2016Estimating,PhysRevLett.119.030402}.
%
For magnetic systems, efforts have been devoted to extract their key parameters from the formation of a spin spiral using ferromagnetic resonance (FMR), Brillouin light scattering (BLS) or neutron scattering (NS). \cite{eyrich2012exchange, Buford2016Estimating, PhysRevLett.119.030402}
%
%
These approaches suffer from the inevitable measurements of time-resolved dynamics and to do so locally.
Imaging by microscope is another important characterization method.
With recent advance in magnetic observing technique, the experimental 
images are able to provide more detailed information of spin configurations.
%
The spin configurations are determined by the magnetic Hamiltonian, however, extracting the exact values of these parameters from merely an image is not an easy task.
What is needed is a method that can estimate the Hamiltonian parameters automatically and appropriately from an experimental observation.

\begin{figure*}[htb]
	\begin{center}
		\includegraphics[width = 12cm]{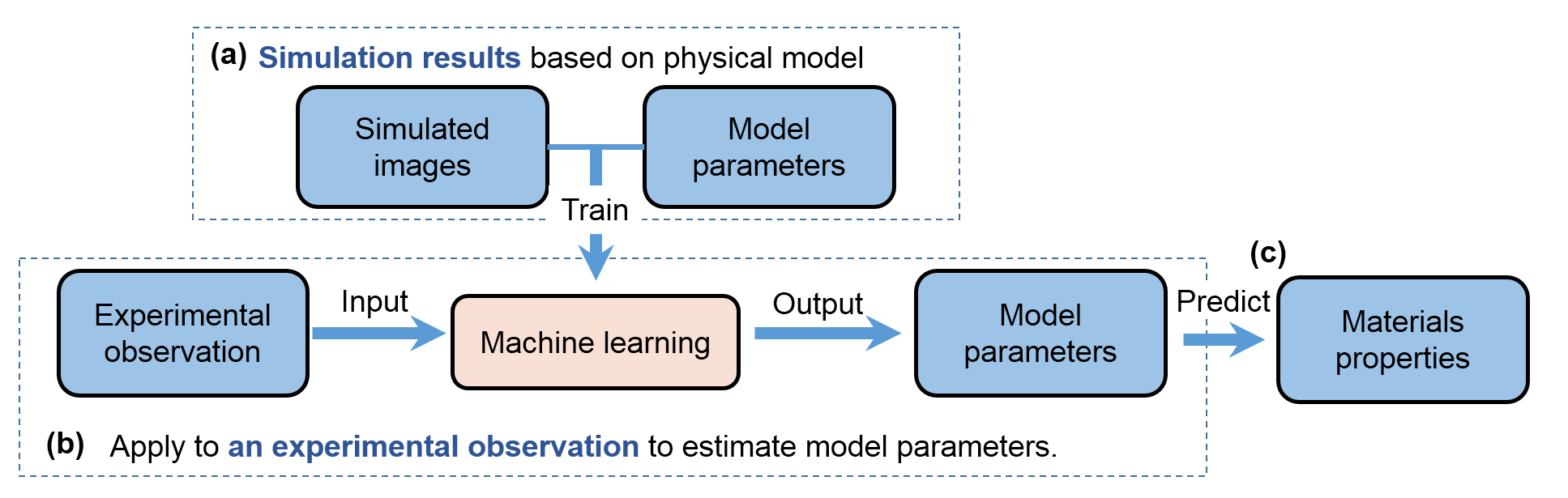}
		\caption{Flow chart of our approach. (a) For a particular experimental observation, a training dataset is formed using simulated images of different Hamiltonian parameters under the same conditions as the experiment. 
%
A machine learning model is trained on these images labeled by the Hamiltonian parameters, and consequently is capable to estimate Hamiltonian parameters from a new image.
(b) An experimental observation is input into the well trained machine learning model, which outputs the corresponding Hamiltonian parameters of that observation.
(c) The estimated Hamiltonian parameters can then be used to predict the properties of material. }
		\label{fig:0}
	\end{center}	 
\end{figure*}

Machine learning (ML) algorithms, such as tree based models \cite{clark2017tree}, kernel based regressors \cite{xue2016accelerated}, and artificial neural networks\cite{Carrasquilla2017Machine,Evert2016Learning,rem2019identifying,zhang2019machine,Carleo2016Solving,bohrdt2019classifying,greplova2017quantum,valenti2019hamiltonian}, learn from the labeled data and predict the unexplored search space, providing a prevalent tool in condensed-matter research.
%
Examples of this include learning the phases and phase transitions of matters \cite{PhysRevX.7.031038,Carrasquilla2017Machine,Evert2016Learning,rem2019identifying,zhang2019machine}, solving the quantum many-body problems \cite{Carleo2016Solving},classifying the snapshots of ultracold atoms \cite{bohrdt2019classifying}, and estimating quantum parameter from quantum measurement \cite{greplova2017quantum,valenti2019hamiltonian}.
Given the success of machine learning in the classification problems in the examples above, the next challenge is to quantitatively extract the physical model parameters from images, especially from experimental ones, so that more abundant information can be obtained.
To do so, the most severe obstacle is the insufficient labeled experimental data to construct a good ML model.
%
\section*{II. FRAMEWORK}
Here we propose an approach based on a combination of numerical simulation and machine learning to achieve parameters estimation from an experimental image .
\autoref{fig:0} shows the workflow chart of our approach. 
For a particular experimental observation, we simulate images using Hamiltonian parameters under the same conditions as the experiment.
A ML model is trained on these images labeled by the Hamiltonian parameters, and consequently is capable to estimate Hamiltonian parameters from a new image.
Then an experimental observation is input into the well trained ML model, which outputs the corresponding Hamiltonian parameters of that observation.
The estimated Hamiltonian parameters can then be used to predict the properties of material.
Our approach allows us to estimate Hamiltonian parameters simultaneously with a single input of experimental observation, without any {\it prior} knowledge.
We demonstrate the success of our approach by precisely estimating three key magnetic parameters ($A_{ex}$, $DMI$ and $M_{sat}$) from input of a spin configuration observation.
The real materials properties such as the hysteresis can then be predicted and validated.
Our work provides a new way to perform parameters estimation in an accelerated, accurate, and efficient manner.
%
%
%
%

%
%
\begin{figure*}[htb]
	\begin{center}
		\includegraphics[width = 15cm]{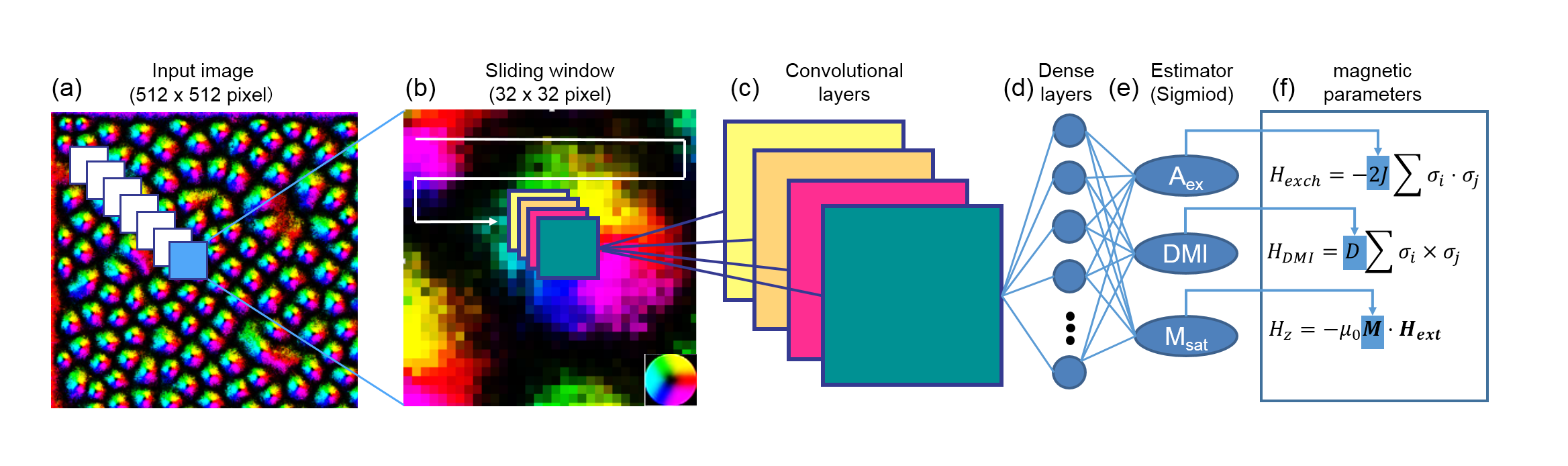}
		\caption{The implementation of a convolutional neural network (\emph{CNN}) to estimate Hamiltonian parameters from images of spin configuration. 
			(a) Input images of the spin configuration (each pixel represents a spin, the color represents the spin orientation) which are labeled with magnetic parameters of $A_{ex}$, $DMI$, and $M_{sat}$. 
			The sliding window on the input images enlarges the training observations. 
			(b) The slided small windows are inputed to a deep convolutional neural network with a variety of layers including 
			(c) convolutional filters, 
			(d) fully connected layers and 
			(e) a output layer. 
			The output layer is set as a  estimator actived by $sigmoid$ to output continuous values of parameters. 
			The three neurons of final sigmoid layer outputs the value of $A_{ex}$, $DMI$, and $M_{sat}$. 
			With those estimated parameters, one can predict materials behaviors and understand underlying physics. }
		\label{fig:1}
	\end{center}	 
\end{figure*}

\begin{figure*}[t]
	\begin{center}
		\includegraphics[width =14cm]{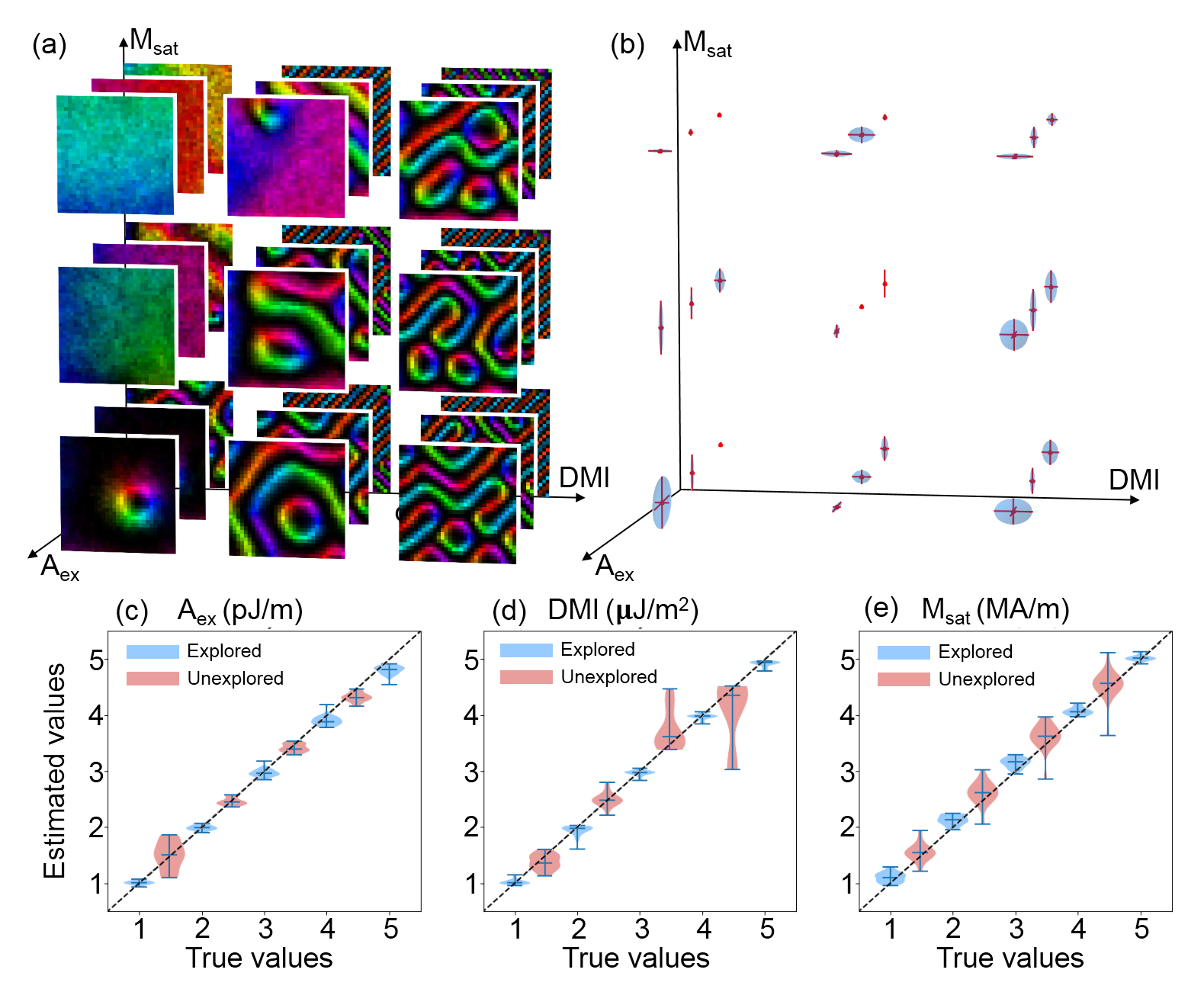}
		\caption{(\textbf{a}) Spin configurations generated by micro-magnetic simulation with the same parameter sets of $A_{ex}$,$DMI$,$M_{sat}$ as the training data but with different initial random seeds. They possess different configurations compared with training data, but contain the same information of parameters. 
			(\textbf{b}) The deviations of estimation for the data of (a) in the parameter space. Error bars in each direction indicate the distance between estimation and the true values of the parameters. 
			(\textbf{c-e}) Plots of the estimated values as the function of true values of the $A_{ex}$,$DMI$,$M_{sat}$ respectively. The blue ones are parameters used in the training data while the red ones represent parameters which are absent in the training data.
		}
		\label{fig:2}
	\end{center}
\end{figure*}
\section*{III. EXPERIMENT}
The Hamiltonian parameters we focus on here are three key parameters ($A_{ex}$, $DMI$ and $M_{sat}$) of a magnetic system. 
We combines micro-magnetic simulation together with a convolutional neural network (\emph{CNN}) to perform the estimation process as shown in \autoref{fig:0}.
%
%

The first stage is the preparation of the training dataset for \emph{CNN}.
The dataset presumably contains images of spin configurations and their corresponding magnetic parameters, as shown in \autoref{fig:1}(a), respectively.
%
%
%
However, collecting such a dataset experimentally still remains challenging as it is prohibitively laborious and expensive.
Borrowed the idea from transfer learning of the robot training\cite{chebotar2019closing}, we generate a training dataset containing simulated spin configurations under a particular temperature and an external field by micro-magnetic model together with the magnetic parameters used.
Robustness of the micro-magnetic model allows us to train a \emph{CNN} that can be adapted to experimental results.
%
To reduce the time cost of the simulation, we simulate monolayer sample to approximate thin film sample or thin specimen used during the transmission electron microscope observation.
But the idea is general and can be extended to more complex situations.
In the second stage, a \emph{CNN} architecture is established, which consists of convolutional layers and dense layers as shown in \autoref{fig:1}(c) and (d), respectively.
%
%
Unlike the traditional \emph{CNN} that directly input the image to the convolutional layers, we introduce a data augmentation method of the sliding window to generate more input data on a small amount of simulated images 
before the convolutional layer, as shown in \autoref{fig:1}(a) to (b).
%
An advantage of the physical system comparing to the inputs of conventional image processing is that the underlying information is distributed evenly, {\it i.e.}, any part of the image contains the same information from one set of model parameters. 
Thus we cut many portions of the input image by sliding the window, which serve as training images with their parameters known.
%
This step greatly enlarges our training dataset and consequently leads to a better \emph{CNN}. 
%
%
Moreover, we replace the last layer of conventional \emph{CNN} (usually a classifier) with an estimator by changing the active function from $softmax$ to $sigmoid$.
Doing so enables the \emph{CNN} to estimate continuous values, as shown in \autoref{fig:1}(e).
Specifically, the estimator layer includes three nodes, and each node will output a value of a particular magnetic parameter of $A_{ex}$, $DMI$ and $M_{sat}$.
By utilizing the trained \emph{CNN}, these parameters for a particular spin configuration can be extracted, and then the prediction of materials properties or the understanding of physics of magnetic phenomena by models such as micro-magnetic simulation can be performed, as shown by \autoref{fig:0}(b) to (c).
The performance of our \emph{CNN} model is then evaluated with simulated test data and real experimental images, respectively. 
As shown in \autoref{fig:2}, our \emph{CNN} model performs well for the testing data set which are generated by micro-magnetic simulation but have not appeared in the training process.
\autoref{fig:2}(a) shows a set of simulated spin configurations that are different from our training data but generated using the same set of magnetic parameters as the training spin configuration.
The random initial seeds are different, thus different spin configurations are obtained.
%
%
%
Their magnetic parameters of $A_{ex}$, $DMI$, and $M_{sat}$ are estimated by our \emph{CNN} model with the inputs in \autoref{fig:2}(a) and are shown in the parameter space in \autoref{fig:2}(b).
%
The size of the ellipsoid indicates the deviation of the estimated values from the true ones.
The estimation of $A_{ex}$ is very good, and deviations of $DMI$ and $M_{sat}$ are slightly larger but still within an acceptable level.
%
The estimated values are then plotted as a function of the true values, as shown in \autoref{fig:2}(c) - (e) by blue violins.
The data distribute along the diagonal line, indicating a precise estimation.
%
%
To test whether our CNN model can estimate parameters that are absent in the training dataset, we generate 4$\times$4$\times$4 new spin configurations.
Both the configurations and parameters are absent in the training dataset.
%
Thus we consider those parameters are unexplored ones.
%
As shown in \autoref{fig:2}(c) - (e) by pink violins, the unexplored parameters are also around the diagonal line, revealing the robustness of our \emph{CNN}. 
%

\begin{figure}[htb]
	\begin{center}
		\includegraphics[width= \columnwidth]{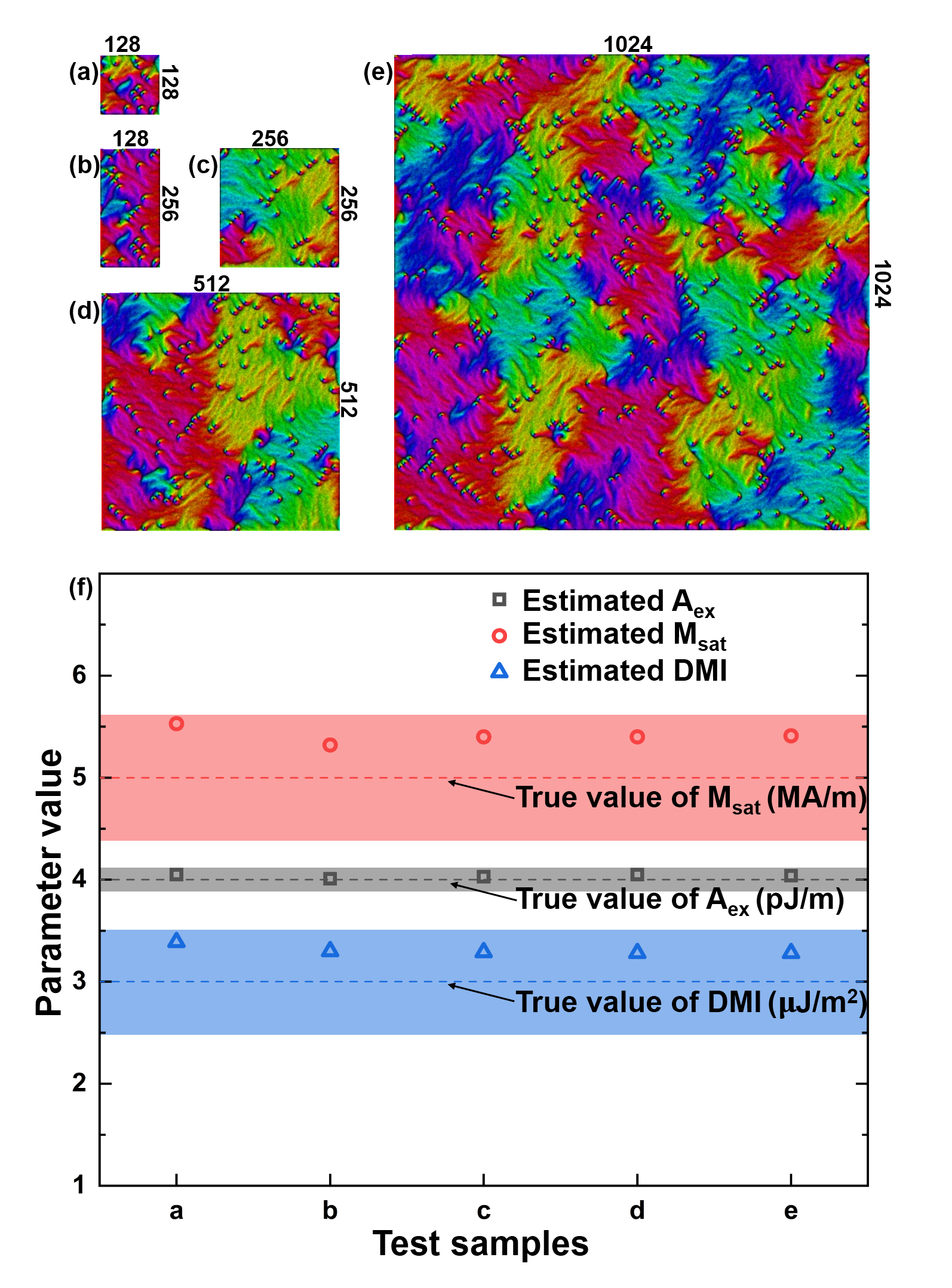}
		\caption{(\textbf{a-e}) Spin configurations from micro-magnetic simulation with the same values of $A_{ex}$, $DMI$, and $M_{sat}$, but different image sizes of 128$\times$128, 128$\times$256, 256$\times$256, 512$\times$512, and 1024$\times$1024.  (\textbf{f}) The estimated parameter values from images of different sizes shown in (\textbf{a-e}). The relative error for $A_{ex}$ is less than 2\%, and that for $DMI$ and $M_{sat}$ is around 10\%.}
			\label{fig:3}
	\end{center}
\end{figure}


%
The above testing result has proved the forecasting capability of our \emph{CNN}, which is achieved by learning patterns from spin configurations rather than similarity measurement or remember the spin configurations.
As the experimental observation varies in size, the generalization ability of our \emph{CNN} to any size of image is of importance.
To validate such an ability, we can estimate the parameters from input images with different sizes rather only the size of $512 \times 512$ used in our training data.
We generate 5 images of different sizes from the same set of parameters as shown in \autoref{fig:3}(a) - (e).
%
%
Noted that the morphology of spin configurations depends on the sample size.\cite{Mulkers2016Cycloidal}
%
\autoref{fig:3}(f) plots the estimated values from different inputs, comparing with the true ones.
It can be seen that the \emph{CNN} performs well regardless of the size of the input image.

\begin{figure}[htb]
	\begin{center}
		\includegraphics[width= \columnwidth]{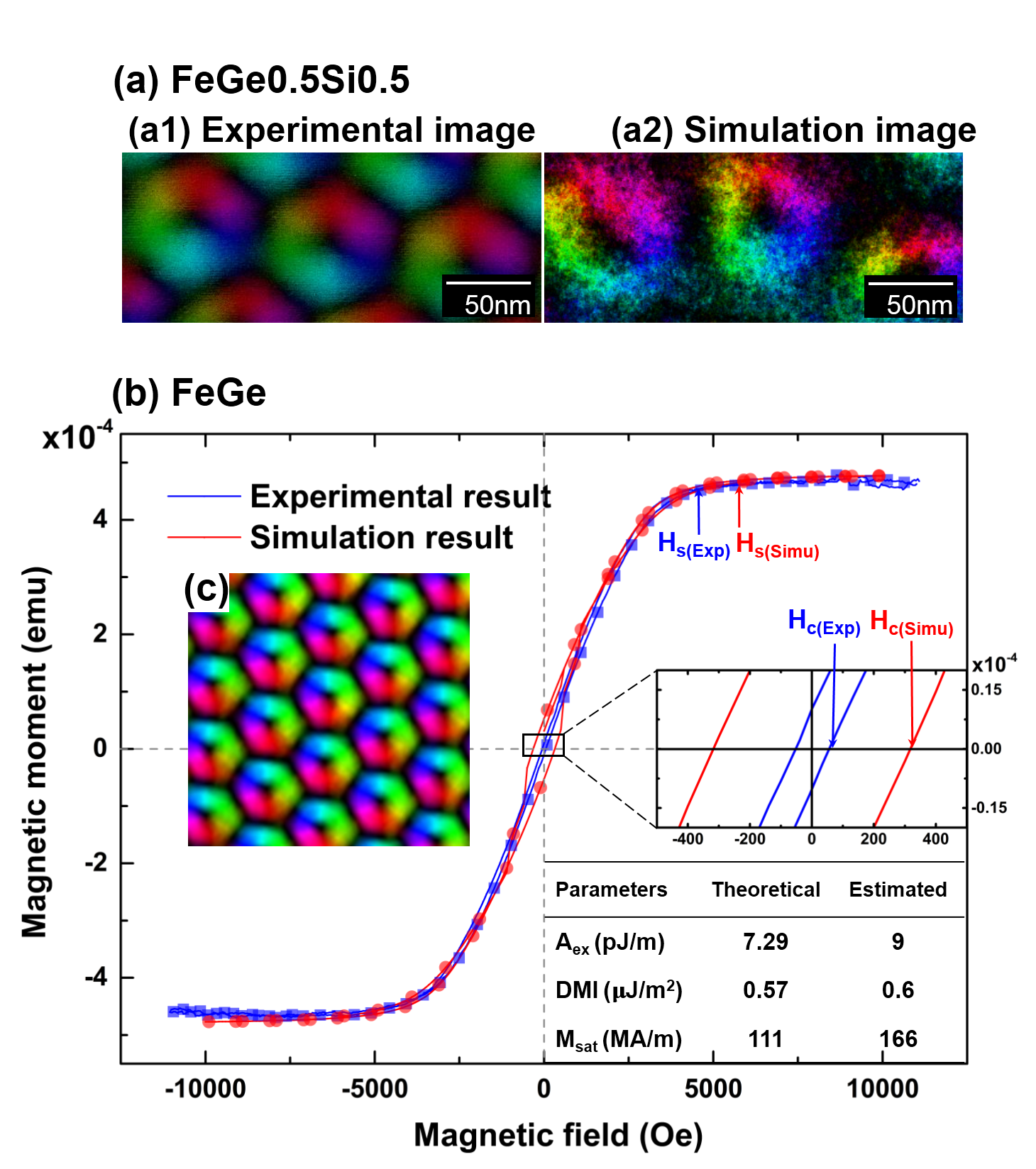}
		\caption{
			(\textbf{a}) The comparison of spin configurations between the experimental image\cite{Matsumoto2016Direct} (input to our \emph{CNN}) and the simulated image using estimated parameters by our \emph{CNN}. 
			(\textbf{b}) Inputing the spin configuration\cite{esser2018high} shown in (\textbf{c}) to our \emph{CNN}, the parameters are estimated and comparable to the theoretical values\cite{PhysRevB.95.220406} as shown in the inset table. 
			The hysteresis loop predicted by the micro-magnetic simulation using these estimated values is in agreement with the experimental one\cite{Huang2012Extended}. 
			The saturation field $H_{s}$ is defined at the knee point in the M-H curve. 
			The coercive field $H_{c}$ is defined at the intersection point between loop and x axis. 
			}
			\label{fig:4}
	\end{center}
\end{figure}

With the power of estimating parameters from input images of various sizes with different parameter sets, a key advantage of our \emph{CNN} is its ability to be directly adapted to real experimental image. 
%
We chose FeGe \cite{esser2018high} and FeGe$_{0.5}$Si$_{0.5}$ \cite{Matsumoto2016Direct} as examples to validate our \emph{CNN} model.
These materials are of great interest due to the existence of the topological phase of skyrmions.
We follow the workflow in \autoref{fig:0} to estimate three intrinsic parameters of $A_{ex}$, $DMI$, and $M_{sat}$.
%
For each case of the two experimental observations, we generate a training dataset utilizing the same temperature and magnetic field used in the experiment to train a \emph{CNN} following the implementation shown in \autoref{fig:1}.
As we have the sliding window step, we do not need to generate a large amount of training data and consequently the process is rather efficient.
%
%
The results of two examples are shown in \autoref{fig:4}.
%
%
An experimental skyrmion lattice of FeGe$_{0.5}$Si$_{0.5}$ specimen by Lorentz transmission electron microscope (TEM) image reconstruction is shown in \autoref{fig:4}(a1).
The observation is performed at 95K under 160mT.
Although the nominal composition of Si is 0.5, the actual composition is hard to determine and thus a precise estimation of parameters of this material is not an easy task.
We generate a training dataset with 5$\times$5$\times$5 spin configurations by micro-magnetic simulation at temperature of 95 K and under magnetic field of 160 mT. 
A \emph{CNN} model is trained on this dataset and estimates the magnetic parameters of $A_{ex}$, $DMI$, and $M_{sat}$ by inputing the experimental skyrmion lattice shown in \autoref{fig:4}(a1).
The spin configuration then can be reproduced from the micro-magnetic simulation using the estimated parameters, and the result is shown in \autoref{fig:4}(a2). 
%
%
The reproduced configuration exhibits very similar features with the experimental observation, indicating a good estimation of these magnetic parameters. 
Besides reproducing the spin configuration, it is also possible to predict the macroscopic properties of a material from only an experimental observation.
A skyrmion lattice of FeGe thin film is shown in \autoref{fig:4}(c). 
It is observed by Lorentz TEM at 265 K under 0.18 T.
Adaptively, we generate 5$\times$5$\times$5 spin configurations by micro-magnetic simulation using temperature of 265 K and magnetic field of 0.18 T and train a new \emph{CNN} to perform the estimation.
As shown in the inset table of \autoref{fig:4}(b), the estimated parameters are in agreement with the theoretical values for this kind of material, which are obtained by microwave absorption spectroscopy. \cite{PhysRevB.95.220406}
We further predict the hysteresis loop of FeGe at the observation temperature 265 K.
%
%
The coercive field ($H_{c}$) and saturation field ($H_{s}$) of the predicted hysteresis loop, which are intensive properties of material, are in agreement with the experimental result \cite{Huang2012Extended} as shown in \autoref{fig:4}(b). 
%
%
%
Since we are not able to get access to the volume of the experimental sample, we vary the sample volume in our simulation to fit the experimental value of the magnetic moment, which is an extensive property.
So that we can estimate the actual volume of the experimental sample around $1mm\times1mm\times3nm$, which is reasonable for a SQUID measurement.
%
Here we employ such an adaptive strategy to train \emph{CNN} that apply to the particular parameters estimation problems.
However, it is possible to include the temperature and magnetic field as the tuning parameters, which requires ``big data'' to train, so that the trained \emph{CNN} can be more general and applied to any  experimental observations directly.

\section*{IV. DISCUSSION}

The mapping between the spin configurations and magnetic parameters are rather complex, for example, there are infinite possible configurations from one set of parameters due to the fluctuations and initial randomness.
Traditionally, the manually designed descriptors to the spin configuration could inevitably loss part of useful information, which makes the estimation of parameters hard.
The \emph{CNN}  which automatically designs as many descriptors of the spin configures as possible and extracts the most relevant features, is so far the best approach to handle this complex mapping problem.
The above validations shows that it is rather possible to acquire magnetic parameters from the spin configuration by a \emph{CNN} machine learning model.

The key ingredients of our approach include: 1) to overcome the shortage of well labeled experimental data, we train a \emph{CNN} on a small training data with images generated by micro-magnetic simulation, which is adaptable to a particular experimental observation with certain condition such as sample shape, temperature, field, and resolution; 2) we propose a data augmentation method: sliding initial image to effectively enlarges the number of input images as the information of parameters distributed evenly throughout the whole spin configuration; and 3) setting the last layer of \emph{CNN} to be an estimator for continuous values instead of the classifier for discrete ones.

%
%
%

\section*{V. CONCLUSION}
In summary, we demonstrate a direct and efficient estimation of magnetic parameters from a single observation of spin configuration {\it via} combination of numerical simulation and machine learning.
%
%
%
%
%
%
%
%
%
Specifically, we demonstrate how to estimate targeted magnetic parameters {\it via} machine learning from only a single experimental image without any other experimental inputs. 
Such an adaptive feature of our approach allows us to deploy it to various experimental observations under different conditions to estimate magnetic parameters, which are usually lack of enough labeled data. 
%
%
The estimated parameters together with numerical simulations based on Hamiltonian of that system can provide many information of the system, such as the micrographies, macroscopic properties, the phase diagram \cite{Yu2010Real} and so on. 
It is thus likely to accelerate the discovery of new materials such as skyrmions with the help of these predictions.
%
%
%
%
%
%
%
Our approach provides a new paradigm to bridge theoretical Hamiltonian to the real material using the combination of numerical simulation and machine learning.
It can be generalized to other condensed matter systems whose microstructure and properties can be described by a Hamiltonian.
%
%
%

\section*{Acknowledgements}
Dingchen Wang and Songrui Wei contribute equally to this work. 
The authors thank Shi Feng and Yifei Tang for the useful discussion and suggestion. 
This research was funded by the National Natural Science Foundation of China (Grants Nos. 51601140, 51701149, 51671157 and 51621063), the Fundamental Research Funds for the Central Universities (China), the World-Class Universities (Disciplines), the National Science Basic Research Plan in the Shaanxi Province of China (2018JM5168), the Characteristic Development Guidance Funds for the Central Universities.

%

\section*{Appendix}
\subsection{Micro-magnetic simulation}
A GPU-accelerated micro-magnetic simulation program, $MuMax^{3}$, generates spin configurations under different parameter sets and with different initial magnetization seeds \cite{Vansteenkiste2014The}.\par
For the studies in \autoref{fig:2}, the environment parameters and geometry parameters have been set as:
\begin{tcolorbox}[colback=white]
	Temp = 300K\\
	B$_{-}$ext = (0,0,0.18)\\\\
    setgridsize(512, 512, 1)\\
    setcellsize(4e-9, 4e-9, 1e-9)
\end{tcolorbox}
\begin{figure}[htb]
	\begin{center}
		\includegraphics[width= \columnwidth]{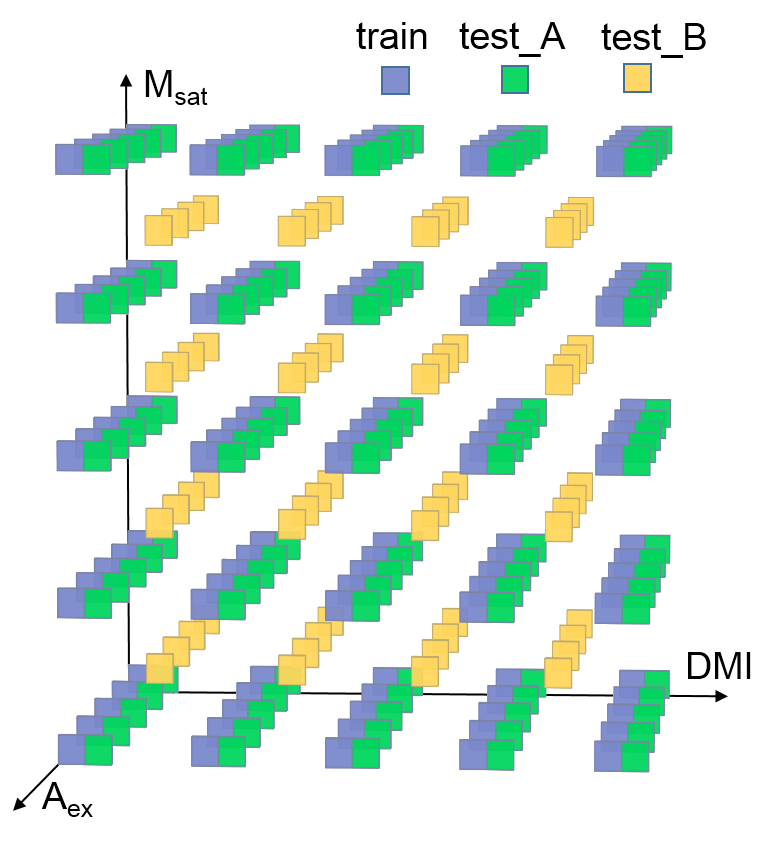}
		\caption{Illustration of our data used. Blue blocks represent training data, which are generated with an initial magnetization seed 1. Green blocks represent test\_A data, which has the same magnetic parameters as the training data but with an initial magnetization seed 2. Yellow blocks represent test\_B data, which are generated from unexplored magnetic parameters.
		}
		\label{fig:5}
	\end{center}
\end{figure}
To generate the training data, magnetic parameters are set as $A_{ex}$=1,2,3,4,5 $(p J / m)$, $DMI$=1,2,3,4,5 $\left(\boldsymbol{\mu} \mathrm{J} / \mathrm{m}^{2}\right)$ and $M_{sat}$=1,2,3,4,5 $(\mathrm{MA} / \mathrm{m})$.
In total, there are 125 parameter sets. 
And 125 spin configurations under these parameter sets with an initial magnetization seed 1 are generated, as shown by blue color label $train$ in \autoref{fig:5}.

In order to test, we first generate test\_A dataset, which includes 125 spin configurations under the parameter sets having been explored before, with an initial magnetization seed 2 as shown by green color label in \autoref{fig:5}. 
Furthermore, we generate test\_B dataset, which contains 64 spin configurations under parameter sets of $A_{ex}$ = (1.5,2.5,3.5,4.5), $DMI$ = (1.5,2.5,3.5,4.5) and $M_{sat}$ = (1.5,2.5,3.5,4.5), which are not explored by the \emph{CNN} yet, as shown by yellow color label in \autoref{fig:5}.

In the experimental image estimation part, the environment parameters and geometry parameters have been set the same as the observation conditions.

\subsection{Convolutional Neural Network}
\begin{table}[htb]
	\caption{Convolutional Neural Network Architecture}
	\label{table:1}
	\begin{tabular}{llll}
		\hline
		\rowcolor[HTML]{80CCFF}
		Layer & \vtop{\hbox{\strut Layer}\hbox{\strut name}} & \vtop{\hbox{\strut Layer}\hbox{\strut function}} & Layer description \\ \hline
		\rowcolor[HTML]{B3E0FF}
		1     & original image & Image input       & \vtop{\hbox{\strut 512*512*3 image}\hbox{\strut of PNG format}}  \\
		\rowcolor[HTML]{99D6FF} 
		2     & \vtop{\hbox{\strut overlapping}\hbox{\strut sliding window}} & Cut image  & \vtop{\hbox{\strut window size:32*32}\hbox{\strut sliding step size:8}}  \\
		\rowcolor[HTML]{B3E0FF}
		3     & conv$_{-}$1    & Convolution       & \vtop{\hbox{\strut 64 3*3*3 convolutions}\hbox{\strut with strides}}        \\
		\rowcolor[HTML]{99D6FF} 
		4     & relu$_{-}$1    & Relu              & \vtop{\hbox{\strut Rectified-linear}\hbox{\strut unit layer}} \\
		\rowcolor[HTML]{B3E0FF}
		5     & padding$_{-}$1 & Padding           & Zero padding      \\
		\rowcolor[HTML]{99D6FF} 
		6     & maxpooling     & Maxpooling        & Maxpooling        \\
		\rowcolor[HTML]{B3E0FF}
		7     & dropout$_{-}$1 & Dropout           & 25$\%$ dropout    \\
		\rowcolor[HTML]{99D6FF} 
		8     & conv$_{-}$2    & Convolution       & \vtop{\hbox{\strut 128 3*3*64 convolutions}\hbox{\strut with strides}}        \\
		\rowcolor[HTML]{B3E0FF}
		9     & relu$_{-}$2    & Relu              & \vtop{\hbox{\strut Rectified-linear}\hbox{\strut unit layer}} \\
		\rowcolor[HTML]{99D6FF} 
		10    & padding$_{-}$2 & Padding           & Zero padding      \\
		\rowcolor[HTML]{B3E0FF}
		11    & dropout$_{-}$2 & Dropout           & 25$\%$ dropout    \\
		\rowcolor[HTML]{99D6FF} 
		12    & fc$_{-}$1      & \vtop{\hbox{\strut Fully}\hbox{\strut connected}} & \vtop{\hbox{\strut fc layer with}\hbox{\strut 512 neurons}} \\
		\rowcolor[HTML]{B3E0FF}
		13    & fc$_{-}$2      & \vtop{\hbox{\strut Fully}\hbox{\strut connected}} & \vtop{\hbox{\strut fc layer with}\hbox{\strut 64 neurons}} \\
		\rowcolor[HTML]{99D6FF} 
		14    & dropout$_{-}$3 & Dropout           & 50$\%$ dropout    \\
		\rowcolor[HTML]{B3E0FF}
		15    & sigmoid        & Sigmoid           & Sigmoid           \\
		\rowcolor[HTML]{99D6FF} 
		16    & estimator      & \vtop{\hbox{\strut Estimation}\hbox{\strut output}}        & MSE Loss           \\
		\hline 
	\end{tabular}
\end{table}
As our input data contains parameters information homogeneously, we employ overlapping sliding window to enlarge our data.
We have studied a variety of data augmentation methods and found that only the sliding window works on the spin configuration. 
Other methods such as scaling and rotation will change the meaning of spin configurations.
We have found the overlapping sliding window method is better than the non-overlapping sliding windows, and the best window size equals to 32 and the best slide step size equals to 8.
Such a setting can help \emph{CNN} perform well while keeping the \emph{CNN} small and easy to train.
Motivated by the success of \emph{CNN} in image recognition, we employ convolutional layers to extract parameters information by feature maps.
We have studied a variety of network architectures and found that convolutional neural networks have much better performance than fully connected networks with the same number of layers.
To achieve the estimation task, we apply the last layer a sigmoid activation function.
The detailed architecture is defined in \autoref{table:1}.

During training, the parameters of the \emph{CNN} are adjusted iteratively to minimize a cost function of mean-square-error (MSE). 
Stochastic gradient descent, along with back propagation, is used for lowering the cost function. 
The training is stopped and all parameters of CNN are set when the MSE saturates.
\subsection{Experimental image}
The experimental image of FeGe \cite{esser2018high} is kindly provided by Dr. Esser and that of FeGe$_{0.5}$Si$_{0.5}$ \cite{Matsumoto2016Direct} is kindly provided by Dr. Matsumoto.
%
FeGe spin configuration is observed at 265 K under 50 mT, and the resolution is 2.34 nm/pixel.
FeGe$_{0.5}$Si$_{0.5}$ one is observed at 95 K under 160 mT, and the resolution is 0.54 nm/pixel.
The experimental hysteresis of FeGe \cite{Huang2012Extended} is kindly provided by Dr. Huang, and it is measured under 250K.

\end{document}